# PERTURBING CONFORMAL TURBULENCE

OMDUTH COCEAL[1] and STEVEN THOMAS[2]
*Department of Physics*
*Queen Mary and Westfield College*
*Mile End Road*
*London E1*
*U.K.*

ABSTRACT

We consider perturbations of the non-unitary minimal model solutions of two-dimensional conformal turbulence proposed by Polyakov. Demanding the absence of non-integrable singularities in the resulting theories leads to constraints on the dimension of the perturbing operator. We give some general solutions of these constraints, illustrating with examples of specific models. We also examine the effect of such perturbations on the Hopf equation and derive the interesting result that the latter is invariant under a certain class of perturbations, to first order in perturbation theory, examples of which are given in specific cases.

[1]email: coceal@qmw.ac.uk

[2]email: st@strings3.ph.qmw.ac.uk

# 1 Introduction

Following the recent proposal by Polyakov [1], that turbulent flow in two spatial dimensions may be understood in terms of certain non-unitary conformal field theories (CFT) (at least in the inviscid limit), there has been renewed interest in such systems. It was quickly realized that Polyakov's results do not single out a unique CFT, rather there appears to be an infinite number of suitable theories [2-5], which even includes the case when boundaries are present that lead to non-vanishing one-point functions [1]. Although the appearance of an infinite number of solutions might be an undesirable feature, the power and elegance of CFTs in this context have led to further investigations and predictions concerning two dimensional turbulence [6].

In this paper we shall consider the situation when any of the non-unitary CFT solutions to 2-dimensional turbulence is perturbed by some primary field of that CFT. Whilst the consequences (see e.g. [7]) of perturbing unitary CFT's in this way have been well documented (perhaps the most notable example being the so called C-theorem of Zamolodchikov [8]), less is known concerning non-unitary CFT's [9]. Furthermore we shall investigate the effects that such perturbations have on the Hopf equations that describe the statistical properties of the inviscid turbulent flow. In particular we show, by considering the first few CFT solutions to turbulence, that there always exists primary fields that leave the Hopf equations invariant.

We begin by reminding the reader of some of the basic formulae underlying Polyakov's solution to 2-dimensional turbulence. The turbulent flow of a fluid with viscosity $\nu$ is governed by the Navier-Stokes equations, which in two spatial dimensions take the form

$$\dot{\omega} + \epsilon_{\alpha\beta}\partial_\alpha\psi\partial_\beta\partial^2\psi = \nu\partial^2\omega \tag{1}$$

where $\psi$ is the stream function, related to the velocity $v$ by $v_\alpha = \epsilon_{\alpha\beta}\partial_\beta\psi$ and to the vorticity $\omega$ by $\omega = \partial^2\psi$ . In the statistical formulation of turbulence one considers correlation functions of relevant quantities (velocities and vorticities). Demanding that we have a stationary probability distribution implies that the correlators $< \omega(x_1)\cdots\omega(x_n) >$ are time-independent. This leads to the inviscid Hopf equation:

$$< \dot{\omega}(x_1)\omega(x_2)\cdots > + < \omega(x_1)\dot{\omega}(x_2)\cdots > + \cdots = 0 \tag{2}$$

Polyakov [1] conjectured that we could interpret these correlators in terms of an effective conformal field theory if all the coordinates $x_i$ are within the so-called inertial range: $a \ll x_i \ll l$. Here $l$ is the infrared cutoff and $a$ is the ultraviolet cutoff, which is determined by the viscosity. Specifically, the stream function $\psi$ is assumed to correspond to a certain primary operator of some CFT. In the limit $\nu \to 0$ (ie $a \to 0$), equation (1) reduces to

$$\dot{\omega} = -\epsilon_{\alpha\beta}\partial_\alpha\psi\partial_\beta\partial^2\psi \tag{3}$$

The right hand side is not well-defined and needs to be regularised by point-splitting. The result is

$$\dot{\omega} \sim |a|^{2\Delta_\phi - 4\Delta_\psi}(L_{-2}\bar{L}^2_{-1} - L^2_{-1}\bar{L}_{-2})\phi \tag{4}$$

where $\phi$ is the minimal dimension operator which appears in the OPE of $\psi$ with itself. The inviscid Hopf equation will only be satisfied if $\dot{\omega}$ vanishes. From the above result we see that there are two possibilities. The first is that the operator $(L_{-2}\bar{L}^2_{-1} - L^2_{-1}\bar{L}_{-2})\phi$ is identically zero. This condition is trivially fulfilled by requiring the corresponding CFT to be degenerate on level two. The simplest such solution is the (2,5) minimal model. The second possibility is that

$$\Delta_\phi > 2\Delta_\psi \tag{5}$$

Thus, any theory with a 'positive defect of dimensions' will solve the Hopf equations. A further constraint introduced by Polyakov [1] is the constant enstrophy flux condition,

$$< \dot{\omega}(r)\omega(0) >= const \tag{6}$$

which leads to the condition

$$\Delta_\psi + \Delta_\phi = -3 \tag{7}$$

Physically, constant enstrophy flux guarantees that conformal field theory correctly describes the enstrophy input at large scales being dissipated at small scales. Polyakov found that these requirements are fulfilled by the (2,21) minimal model. Subsequently, Lowe [2] and others [3-5] have shown that there are an infinite number of minimal model solutions of these constraints, the (2,21) model being the simplest one. Lowe also considers an alternative constraint, the constant energy flux condition, which yields

$$\Delta_\psi + \Delta_\phi = -2 \tag{8}$$

This also has an infinite number of solutions. The constant energy flux condition allows for the possibility of an inertial range in which energy cascades from small to large scales as envisaged by Kraichnan [10], Leith [11] and Batchelor [12].

One of the most important physical predictions of conformal turbulence is the exponent of the energy spectrum. This is found to be given by

$$E(k) \sim k^{4\Delta_\psi + 1} \tag{9}$$

Now, since the dimension of $\psi$ is model-dependent, the predicted exponent is not unique. The experimental value lies between 3 and 4, but there are many models which give this result. Some additional constraints are needed to identify the correct theory of conformal turbulence. This has not been done so far.

The simplest solutions are minimal models. In general they will always be non-unitary. This means that there will be some primary operators with negative dimensions in the theory. To see this we note that at least the stream function operator $\psi$ will always have negative dimension. This is because equations (5) and (7) taken together imply that $\Delta_\psi < -1$ while (5) with (8) give $\Delta_\psi < -\frac{2}{3}$. In unitary models the unit operator, with dimension zero, is always the lowest dimension operator. However, for the non-unitary models, due to the negative dimensions, this is no longer the case. Some pertinent facts about minimal models are summarised here.

The $(p,q)$ minimal model (where $p$ and $q$ are co-prime positive integers, with $p < q$) contains $\frac{1}{2}(p-1)(q-1)$ degenerate primary operators $\psi_{m,n}$, where $1 \leq m < p$ and $1 \leq n < q$, with conformal dimensions given by

$$\Delta_{m,n} = \frac{(pn - qm)^2 - (p-q)^2}{4pq} \tag{10}$$

These primary operators satisfy the following fusion rules:

$$\psi_{m_1,n_1} \times \psi_{m_2,n_2} = \sum_i \sum_j D^{(i,j)}_{(m_1,n_1)(m_2,n_2)} [\psi_{i,j}] \ , \tag{11}$$

where $i$ runs from $|m_1 - m_2| + 1$ to $\mathrm{Min}(m_1 + m_2 - 1, 2p - m_1 - m_2 - 1)$, and is odd if $m_1 + m_2$ is even and vice-versa, and similarly for $j$. The $D^{(i,j)}_{(m_1,n_1)(m_2,n_2)}$ are the structure constants of the CFT. The symbol $[\psi_{i,j}]$ denotes the conformal family of $\psi_{i,j}$. The operators $\psi_{r,s}$ and $\psi_{p-r,q-s}$ have the same dimensions, and they are identified. It is found that an infinite number of these minimal models satisfy the constraints (5) and (7) or (8), the simplest being the (2,21) model with $\psi = \psi_{1,4}$. Some results are listed in the paper by Lowe [2].

## 2 Perturbing the solutions of conformal turbulence

We now wish to investigate the effect of perturbing such a (non-unitary) CFT by some primary operator in the theory. In particular we are interested in finding out whether the perturbed theory still satisfies the Hopf equations. This will give some indication on the 'stability' of these solutions.

We proceed in a similar manner described by Cardy [7]. Consider the effect on the 'action' describing the CFT of perturbing by a certain primary operator $\Phi$ with scaling dimensions $(h, h)$:

$$S = S^{*} - \lambda \int \Phi(z, \bar{z}) d^2 z \ , \tag{12}$$

where $S^{*}$ is the fixed point action and $\lambda$ is a coupling constant with conformal dimensions $(1-h, 1-h)$. Evaluating the correlation functions in the Hopf equation (2) perturbatively in $\lambda$, we obtain (in complex notation)

$$\begin{aligned} &< \dot{\omega}(z_1, \bar{z_1})\omega(z_2, \bar{z}_2) \cdots >_S \, = \, < \dot{\omega}(z_1, \bar{z_1})\omega(z_2, \bar{z}_2) \cdots >_{S^*} \\ &+\lambda \int < \Phi(z, \bar{z})\dot{\omega}(z_1, \bar{z_1})\omega(z_2, \bar{z}_2) \cdots >_{S^*} \ d^2 z \ + O(\lambda^2) \end{aligned} \tag{13}$$

The first term is simply the unperturbed Polyakov correlator evaluated in the minimal model. Thus, by (4), it vanishes in the usual Polyakov ansatz, namely equation (5). Assuming this condition, we want to determine the effect of the second term. It is clear that the only contributions which do not obviously vanish as a result of equation (5), are those obtained by evaluating the integral

$$I = \lambda \int \Phi(z, \bar{z}) \ \dot{\omega} \ d^2 z \tag{14}$$

We can use equation (3) to substitute for $\dot{\omega}$, obtaining

$$I = -\lambda \int d^2\omega \ \Phi(\omega, \bar{\omega}) \ [\epsilon_{\alpha\beta}\partial_\alpha \psi(z, \bar{z})\partial_\beta \partial^2 \psi(z, \bar{z})] \ , \tag{15}$$

Equation (15) represents the dressing of the inertial term by the perturbing operator, and as before, the term in brackets is to be understood in terms of its point-splitted definition

$$\epsilon_{\alpha\beta}\partial_\alpha \psi(z, \bar{z})\partial_\beta \partial^2 \psi(z, \bar{z}) = \overline{\lim_{a \to 0}} \epsilon_{\alpha\beta}\partial_\alpha \psi(z + a, \bar{z} + \bar{a})\partial_\beta \partial^2 \psi(z, \bar{z}) \tag{16}$$

where $-$ in $\overline{\lim_{a\to 0}}$ denotes spatial averaging over the cutoff $a$, i.e. if we express $a = |a|e^{i\theta_a}$ then one must integrate over $\theta_a$ before taking the limit $a \to 0$ Substituting this, we then obtain

$$I = \overline{\lim_{a\to 0}} \int d^2\omega \; O(z,\bar z,z',\bar z') \; \psi(z,\bar z)\psi(z',\bar z')\Phi(\omega,\bar\omega) \;, \tag{17}$$

where

$$O(z,\bar z,z',\bar z') \;=\; \partial_{z'}\partial_{\bar z}^2\partial_z \;-\; \partial_{\bar z'}\partial_z^2\partial_{\bar z}$$

and $z' = z+a,\ \bar z' = \bar z+\bar a$. As it stands this integral is ultraviolet divergent and needs to be regularised. Indeed it is because of this that we expect to obtain generally different kinds of operators than the one found by Polyakov in equation (4) . Following [7], we do this by inserting step functions $\Theta[(z-\omega)(\bar z-\bar\omega)-b^2]$ and $\Theta[(z'-\omega)(\bar z'-\bar\omega)-b^2]$ into the integral, (and again taking spatial averaging over the cutoff $b$ ). Next we pull out a $\partial_{\bar z}$ and act with it on the first $\Theta$-function. This gives a $\delta$-function, which enables us to perform the radial integration. The result is

$$I_b = \lim_{a\to 0} \int_0^{2\pi} d\theta_a \int_0^{2\pi} d\theta_b \; |b| \; e^{i\theta_b} \; \partial_z(\partial_{z'}\partial_{\bar z} - \partial_{\bar z'}\partial_z) \; \psi(z',\bar z')[\psi(z,\bar z)\Phi(\omega,\bar\omega)] \;, \tag{18}$$

where $b = |b|e^{i\theta_b}$, $I_b$ is the regularized integral $I$ and the square brackets mean that we have to take the OPE of the last two operators first. Performing this OPE, we obtain

$$\psi(z,\bar z)\Phi(\omega,\bar\omega) \sim \sum_i \sum_j |b|^{2(\Delta_{i,j}-\Delta\psi-\Delta\Phi)}[\psi_{i,j}(z,\bar z)] \tag{19}$$

The limit $a \to 0$ then forces us to take the next OPE. Doing this we find

$$\psi(z',\bar z')[\psi(z,\bar z)\Phi(\omega,\bar\omega)] \;\sim\; \sum_{i,j,k,l} |b|^{2(\Delta_{i,j}-\Delta\psi-\Delta\Phi)}|a|^{2(\Delta_{k,l}-\Delta_{i,j}-\Delta_\psi)} \; [\psi_{k,l}(z,\bar z)] \tag{20}$$

where, following usual convention, $[\psi_{k,l}]$ denotes the conformal class of $\psi_{k,l}$. The descendants of the primary fields are crucial. As in Polyakov's treatment, we need to go beyond the leading term to obtain a non-vanishing result. It is clear from parity considerations that the effect of the operator $(\partial_{z'}\partial_{\bar z} - \partial_{\bar z'}\partial_z)$ in (18) is to yield a lowest level term of the form $(a\bar b - b\bar a)\ D\psi_{k,l}$, where $D$ is a dimension (2,2) parity even operator such as $L_{-2}\bar L_{-2}$, $L_{-1}^2\bar L_{-1}^2$, $L_{-2}\bar L_{-1}^2$ etc. Equation (19) implies that, due to the angular integration, a non-vanishing result will be obtained only if we can cancel the factor of $e^{i\theta_b}$. This means that we need an extra $\bar b$. Hence the only terms that survive the action of the final $\partial_z$ are of the form $\bar{b}D\psi_{k,l}$. Performing the integrations, we finally obtain for the integral $I$

$$I = (2\pi)^2 \lim_{a\to 0, b\to 0} \lambda \sum_{i,j,k,l} |b|^{2(\Delta_{i,j}-\Delta_\psi-\Delta_\Phi+1)} \, |a|^{2(\Delta_{k,l}-\Delta_{i,j}-\Delta_\psi)} \, D \, \psi_{k,l}(z,\bar{z}) \tag{21}$$

To avoid ambiguities in the ordering of the limits $a \to 0$ and $b \to 0$, we shall take both limits simultaneously. The effect of taking the simultaneous limits is that $a$ and $b$ can be considered to be essentially the same cut-off parameter, and therefore we can write

$$I \sim \lim_{a\to 0} \lambda \sum_{k,l} |a|^{2(\Delta_{k,l}-\Delta_\Phi-2\Delta_\psi+1)} \, D \, \psi_{k,l}(z,\bar{z}) \tag{22}$$

Note that the dimensions of the intermediate fields $\psi_{i,j}$ drop from the final result. The exponent of $|a|$ is crucial. To avoid introducing singularities the perturbing operator $\Phi$ has to be such that none of the fields $\psi_{k,l}$ gives rise to a negative exponent. In other words, for any sensible perturbation we have the necessary condition

$$\Delta_{k,l} - \Delta_\Phi - 2\Delta_\psi + 1 \geq 0 \tag{23}$$

The limit $a \to 0$ means that we need only keep the term with the smallest exponent of $|a|$. Let us denote the corresponding lowest dimension field $\psi_{k,l}$ by $\Psi$. We can then write

$$I \sim |a|^{2(\Delta_\Psi-\Delta_\Phi-2\Delta_\psi+1)} \, D \, \Psi \tag{24}$$

This is the direct analogue of equation (4). Here $I$, as defined by (14), is $\dot{\omega}$ dressed by the conformal perturbation . $D$ is a (2,2) parity even operator made up from $L_{-1}$, $\bar{L}_{-1}$, $L_{-2}$ and $\bar{L}_{-2}$, and $\Psi$ is the minimal dimension operator appearing in the OPE $\psi\psi\Phi$, analogous to $\phi$ of equation (4).

We thus find that, in the limit $a \to 0$, $I$ will vanish for perturbations such that

$$\Delta_\Psi - \Delta_\Phi - 2\Delta_\psi + 1 > 0 \tag{25}$$

This is analogous to condition (5). In that case, we conclude from equation (13) that the Hopf equation will still be satisfied, at least to first order in $\lambda$. Hence, the minimal model solutions of conformal turbulence are stable to such perturbations. In other words, although perturbing these CFT's generally breaks their conformal invariance (ie they are no longer CFT's) they still remain solutions of the Navier-Stokes equations.

As emphasised in [7], one has to consider the possibility that new ultraviolet divergences can arise in correlation functions of the perturbing operator. To see how to avoid such divergences, consider correlation functions of the perturbing operator $\Phi$ in the perturbed theory. Similar to (13) we obtain

$$< \Phi(z_1,\bar{z_1})\Phi(z_2,\bar{z_2})\cdots >_S = < \Phi(z_1,\bar{z_1})\Phi(z_2,\bar{z_2})\cdots >_{S^*}$$
$$+\lambda \int < \Phi(z,\bar{z})\Phi(z_1,\bar{z_1})\Phi(z_2,\bar{z_2})\cdots >_{S^*} \; d^2z \; + O(\lambda^2) \tag{26}$$

The ultraviolet behaviour of the integrand in the first order term is determined by the operator product expansion of $\Phi$ with itself

$$\Phi \; \Phi \;\sim\; |a|^{2(\Delta_\varphi - 2\Delta_\Phi)} \; \varphi \tag{27}$$

where $\varphi$ is the minimal dimension operator in the OPE in equation (27). There will be no new ultraviolet divergences provided

$$\Delta_\varphi - 2\Delta_\Phi + 2 > 0 \tag{28}$$

Now, in a unitary theory, since all the conformal dimensions are positive, it is clear that this condition will automatically be satisfied if $\Delta_\Phi \le 1$, as pointed out in [7] i.e. $\Phi$ is a so called relevant perturbation. However, in non-unitary theories the situation is further complicated by the presence of fields with negative dimensions, and more care is required. Our immediate task is to find general solutions of (28). Later we also look for sub-classes of these that satisfy (25), thus leaving the Hopf equation invariant to O($\lambda^2$).

For illustration, we first consider some special cases. For example, if we perturb with the stream function operator $\psi$ itself we have $\Phi \equiv \psi$ and $\varphi \equiv \phi$. Then we can make use of either (7) or (8) to eliminate $\Delta_\varphi$ from (28). The result is $\Delta_\psi < -\frac{1}{3}$ for the constant enstrophy case and $\Delta_\psi < 0$ for the constant energy case. We recall that the corresponding conditions imposed on $\psi$ by conformal turbulence are $\Delta_\psi < -1$, and $\Delta_\psi < -\frac{2}{3}$ in each of these cases respectively. Hence, we can perturb *any* solution of conformal turbulence with the stream function operator $\psi$ without introducing ultraviolet divergences in the 2-point functions of $\psi$.

Let us now be more general. We consider separate cases. For positive $\Delta_\Phi$, we note that, since the unit operator always occurs in the OPE $\Phi$ $\Phi$, the maximum possible value of the first term $\Delta_\varphi$ is 0. Thus, if $\Delta_\Phi > 1$ the second term will dominate and yield an overall negative result, violating (28). Note that this latter result is quite general and in fact applies to any CFT, not just to those satisfying conformal turbulence. The remaining region $0 < \Delta_\Phi < 1$ is more delicate to analyse. It appears that the result is dependent on the particular model under consideration and that no general statement can be made here. We illustrate with the example of the (3,25) model. In this model, we have two such operators: $\psi_{1,16}$ and $\psi_{1,17}$, with dimensions 0.15 and 0.64 respectively (see Table 1). First let $\Phi \equiv \psi_{1,16}$. The minimal dimension field in its OPE is $\psi_{1,9}$, with dimension $\Delta_\varphi = -1.6$. Evaluating the lhs of (28) we find that the inequality is indeed satisfied. Next we choose $\Phi \equiv \psi_{1,17}$. This again gives $\varphi \equiv \psi_{1,9}$. Inserting the values for the corresponding dimensions, we see that (28) is now violated. Thus we obtain opposite results for these two operators.

We next consider the final case where $\Delta_\Phi < 0$. Consider the minimal dimension $\Delta_0$ appearing in the non-unitary minimal $(p, q)$ model. This is given by

$$\Delta_0 = \frac{1-(q-p)^2}{4pq} \tag{29}$$

Defining $R = p/q$ and rearranging equation (29) we obtain

$$\Delta_0 = \frac{1}{4}\left[2 - R - \frac{1}{R}(1 - 1/q^2)\right] \tag{30}$$

Now the smallest value of $q$ occuring in the solutions of conformal turbulence is 21, making the $1/q^2$ term negligible. Thus,

$$\Delta_0 \approx \frac{1}{4}\left[2 - R - 1/R\right] \tag{31}$$

The percentage error introduced by using the latter expression is less than 0.3% for the (2,21) model and decreases rapidly with increasing $q$ for subsequent models. Since $p < q$, the physical domain of $R$ is (0,1). In this range $\Delta_0$ is a monotonically increasing function as illustrated in fig 1. For large $q$, the limit $R = 1$ corresponds to unitary minimal models, for which $q = p + 1$. There we see that $\Delta_0 = 0$, ie the minimal dimension operator in such models is the unit operator, with conformal dimension 0. In other words, no negative dimensions appear in such theories. As far as the minimal models of conformal turbulence are concerned, we saw in the introduction that they are always of necessity non-unitary. For these, Lowe [2] derived some inequalities that $R$ should satisfy. For the constant enstrophy case, these are

$$\frac{1}{10} \; < \; R \; < \; 4 - \sqrt{15} \tag{32}$$

for all solutions except (2,21), for which the lower limit is $\frac{2}{21}$. For the constant energy case, the corresponding range is

$$\frac{3}{22} \; < \; R \; < \; 3 - \sqrt{8} \tag{33}$$

We have computed the values of $\Delta_0$ for these limits, and the corresponding ranges are given by, respectively,

$$-2.025 \; < \; \Delta_0 \; < \; -1.5 \tag{34}$$

(with the lower limit as -2.14 for the (2,21) model) and

$$-1.37 \; < \; \Delta_0 \; < \; -1 \tag{35}$$

Now we go back to equation (28). For negative $\Delta_\Phi$ the second term is positive. If for a moment we take $\varphi$ to be the absolute minimal operator in the model then we note that (28) will automatically be satisfied if we can fulfill the stronger constraint

$$\Delta_0 + 2 \; > \; 0 \tag{36}$$

We note that this is manifestly satisfied by all the solutions of the constant energy constraint. For the constant enstrophy case, the only possible violations arise from the (2,21) model, where $\Delta_0 = 2.14$, and those models with $-2.025 < \; \Delta_0 \; < \; -2$, ie $0.1 \; < \; R \; < \; 0.101$. In each of these, if $\varphi$ is any operator with dimension $< \; -2$, then (28) will be violated if there is any perturbing operator $\Phi$ which contains $\varphi$ in its OPE $\Phi \; \Phi$ such that

$$\Delta_\Phi > 1 + \frac{1}{2}\Delta_\varphi \tag{37}$$

Explicit calculation for the (2,21) model shows that this is not the case. In fact, this last condition seems to be very unlikely to be met by any minimal model. All the models which we have tested seem to indicate that (28) is satisfied anyway, regardless of whether $\Delta_0 < -2$. Unfortunately we have not as yet been able to come up with a rigorous proof to that effect.

To summarise, we have found that all primary operators with negative dimension and some of those with positive dimension $< 1$ can be used to perturb a general non-unitary CFT without introducing ultraviolet divergences.

With regard to the solutions of conformal turbulence, we now return to equation (25) and try and verify if there is at least a subset of these operators that also leaves the Hopf equations invariant to $O(\lambda^2)$.

So far, our treatment has been completely general. All of the above analysis applies to *any* minimal model solution of conformal turbulence. To proceed further we need to examine specific models.

Let us briefly summarise what we want to find. Starting from a minimal model $(p, q)$ satisfying Polyakov's constraints (5) and (7) or (8), with a certain 'stream function' operator $\psi$, we choose different perturbing primary operators $\Phi$ from the model. Denoting by $\Psi$ the minimal dimension operator appearing in the OPE $\psi\psi\Phi$, we look for the $\Phi$'s that will satisfy condition (25). To do this we rearrange (25) to give $\Delta_\Phi < 1 + \Delta_\Psi - 2\Delta_\psi$. Let us now define $k = 1 + \Delta_\Psi - 2\Delta_\psi$. We then calculate the value of $k$ for different choices of $\Phi$. We have done this for the first few models for both the constant enstrophy and the constant energy case. In general, the value of $k$ depends not only on the model but also on the perturbation operator. Although it is not constant for a given model it varies within a well-defined narrow range. We can then easily identify those operators with dimension less than $k$. For example the value of $k$ for the (3,25) model is either 2.19 or 2.2. We find that the fields $\psi_{1,1}, \ldots, \psi_{1,19}$ have dimension less than this (see Table 1). We have investigated the first few low-lying models. The detailed results are shown in Table 2. The value of $k$ lies roughly between 1 and 3. Hence we can conclude that, at least for the models investigated, all the solutions of (28) also satisfy (25), an example being the case $\Phi \equiv \psi$ mentioned above. The converse statement is not necessarily true.

In conclusion, it would certainly be worthwhile extending the approach in this paper to find all possible perturbations that leave the inviscid Hopf equation invariant, at least to leading order in perturbation theory. Since such perturbations also break conformal symmetry, one can view our results as widening the possible solutions of two-dimensional turbulence. When conformal symmetry is broken in this way, one natural question is whether the parameters in the models (e.g. $\lambda$ , the scaling dimensions, or even the viscosity $\nu$ ) undergo renormalization group flow, and if so what might be the fixed points? A related outstanding problem is to give a physical interpretation to the various perturbing operators discussed in this paper. The results presented in [6] could provide a starting point for such an investigation. Such an investigation could place additional constraints on the allowed set of perturbing operators listed in Table 2, for example that $\Delta_\Phi > 0$.

## Acknowledgements

O.C. gratefully thanks the University of London for sponsoring part of this research in the form of a postgraduate studentship, and also acknowledges the help given to him by C.Kobdaj. S.T. would like to thank O. A. Soloviev for useful discussions.

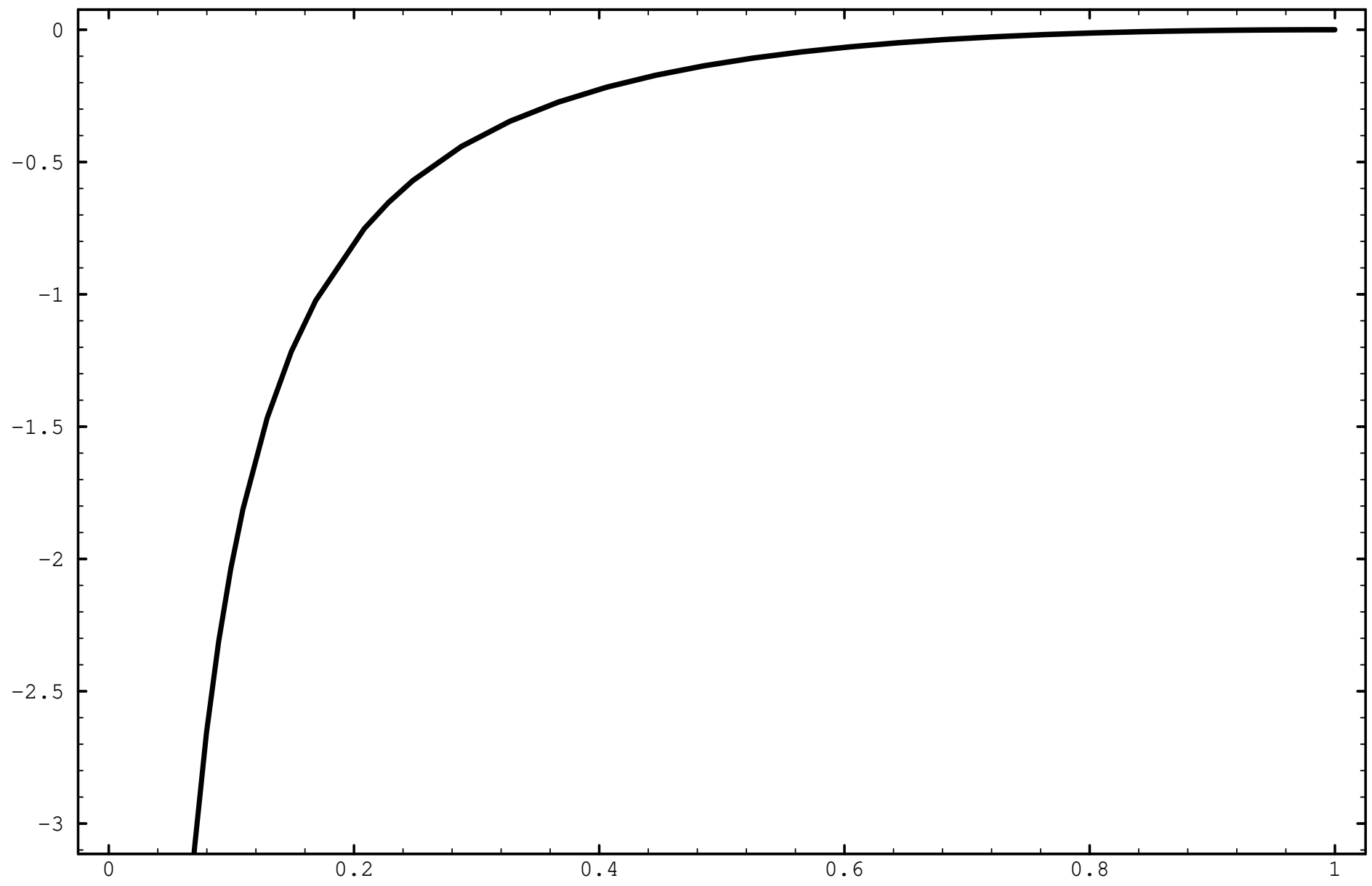

$\Delta_0$

$R$

**Fig 1**

| $\psi_{m,n}$ | $\Delta_{m,n}$ | $\psi_{m,n}$ | $\Delta_{m,n}$ | $\psi_{m,n}$ | $\Delta_{m,n}$ |
|---|---|---|---|---|---|
| $\psi_{1,1}$ | 0 | $\psi_{1,9}$ | -1.60 | $\psi_{1,17}$ | .64 |
| $\psi_{1,2}$ | -.41 | $\psi_{1,10}$ | -1.53 | $\psi_{1,18}$ | 1.19 |
| $\psi_{1,3}$ | -.76 | $\psi_{1,11}$ | -1.4 | $\psi_{1,19}$ | 1.8 |
| $\psi_{1,4}$ | -1.05 | $\psi_{1,12}$ | -1.21 | $\psi_{1,20}$ | 2.47 |
| $\psi_{1,5}$ | -1.28 | $\psi_{1,13}$ | -.96 | $\psi_{1,21}$ | 3.2 |
| $\psi_{1,6}$ | -1.45 | $\psi_{1,14}$ | -.65 | $\psi_{1,22}$ | 3.99 |
| $\psi_{1,7}$ | -1.56 | $\psi_{1,15}$ | -.28 | $\psi_{1,23}$ | 4.84 |
| $\psi_{1,8}$ | -1.61 | $\psi_{1,16}$ | .15 | $\psi_{1,24}$ | 5.75 |

Table 1: Dimensions of primary operators in the (3,25) model.

| $(p,q)$ | $\psi$ | $\Phi$ | $(p,q)$ | $\psi$ | $\Phi$ |
|---|---|---|---|---|---|
| $(2,21)$ | $\psi_{1,4}$ | $\psi_{1,1}\ldots\psi_{1,10}$ | $(7,62)$ | $\psi_{1,13}$ | $\psi_{1,1}\ldots\psi_{1,20}$ |
| $(3,25)$ | $\psi_{1,11}$ | $\psi_{1,1}\ldots\psi_{1,19}$ | | | $\psi_{2,7}\ldots\psi_{2,28}$ |
| $(3,26)$ | $\psi_{1,5}$ | $\psi_{1,1}\ldots\psi_{1,20}$ | | | $\psi_{3,16}\ldots\psi_{3,37}$ |
| $(5,32)$ | $\psi_{1,9}$ | $\psi_{1,1}\ldots\psi_{1,14}$ | $(8,47)$ | $\psi_{3,17}$ | $\psi_{1,1}\ldots\psi_{1,14}$ |
| | | $\psi_{2,5}\ldots\psi_{2,21}$ | | | $\psi_{2,4}\ldots\psi_{2,20}$ |
| $(6,37)$ | $\psi_{1,8}$ | $\psi_{1,1}\ldots\psi_{1,14}$ | | | $\psi_{3,10}\ldots\psi_{3,26}$ |
| | | $\psi_{2,4}\ldots\psi_{2,20}$ | | | $\psi_{4,16}\ldots\psi_{4,23}$ |
| | | $\psi_{3,11}\ldots\psi_{3,18}$ | $(8,67)$ | $\psi_{3,28}$ | $\psi_{1,1}\ldots\psi_{1,19}$ |
| $(6,55)$ | $\psi_{1,14}$ | $\psi_{1,1}\ldots\psi_{1,20}$ | | | $\psi_{2,6}\ldots\psi_{2,27}$ |
| | | $\psi_{2,8}\ldots\psi_{2,29}$ | | | $\psi_{3,14}\ldots\psi_{3,36}$ |
| | | $\psi_{3,17}\ldots\psi_{3,27}$ | | | $\psi_{4,23}\ldots\psi_{4,33}$ |

Table 2: Allowed perturbations of the first nine minimal models.